\documentclass[%
 reprint,
 amsmath,amssymb,
 aps,
]{revtex4-2}

\usepackage{graphicx}% Include figure files
\usepackage{dcolumn}% Align table columns on decimal point
\usepackage{bm}% bold math
\usepackage{dcolumn}
\usepackage{bm}
\usepackage{graphicx}
\usepackage{epstopdf}
\usepackage{subfigure}
\usepackage{tikz}
\usepackage{siunitx}
\usepackage{float}
\usepackage{booktabs}
\usepackage{url}
\usepackage[utf8]{inputenc}
\usepackage{color}
\usepackage{makecell}
\usepackage{threeparttable}
\usepackage{tikz}
\usepackage[compat=1.1.0]{tikz-feynman}
\usepackage{tikz-feynman}
\usepackage{geometry}
\usepackage{slashed}

\usetikzlibrary{shapes,arrows}

\begin{document}

%\preprint{APS/123-QED}

\title{Light baryon spatial correlators at short distances}% Force line breaks with \\

\author{Chao Han}
\email{chaohan@sjtu.edu.cn}
\author{Jialu Zhang}%
 \email{elpsycongr00@sjtu.edu.cn}
\affiliation{INPAC, Key Laboratory for Particle Astrophysics and Cosmology (MOE), 
Shanghai Key Laboratory for Particle Physics and Cosmology, School of Physics and Astronomy, Shanghai Jiao Tong University, Shanghai 200240, China
}%

%\collaboration{M\widetilde{O}_{b}O Collaboration}%\noaffiliation

%\affiliation{
%\affiliation{
%\affiliation{%
 %This line break forced with %\textbackslash\textbackslash
%}%

%\collaboration{}%\noaffiliation

\date{\today}% It is always \today, today,

\begin{abstract} 

To study the light baryon light-cone distribution amplitudes (LCDAs), the spatial correlator of the light baryon has been calculated up to one-loop order in coordinate space. 
They reveal certain identities that do not appear in the study of mesons DAs and PDFs.
Subsequently, it was renormalized using the ratio renormalization scheme involving division by a 0-momentum matrix element. 
Then through the matching in the coordinate space, the light baryon light-cone correlator can be extracted out from the spatial correlator.
%By choosing a specific type of spatial correlator for light baryons, the LCDA is determined from first principles.
These results provide insights into both the ultraviolet (UV) and infrared (IR) structures of the light baryon spatial correlator, which is valuable for further exploration in this field. 
Furthermore, the employed ratio scheme is an efficient and lattice-friendly renormalization approach suitable for short-distance applications. 
These can be used for studying the light baryon LCDAs using lattice techniques.
\end{abstract}

\maketitle

%\tableofcontents

\section{introduction}

Light-cone distribution amplitudes (LCDAs) for light baryons serve as fundamental components in the description of these light baryons.
They are defined through the QCD factorization for the exclusive process with a large momentum transfer, and encodes the crucial non-perturbative physics within the light baryons \cite{Shih:1998pb}.
%These LCDAs are defined within the framework of QCD factorization, specifically for exclusive processes involving significant momentum transfers. 
They encapsulate vital non-perturbative information inherent to light baryons.
These distribution amplitudes hold a key position in unraveling the inner structures of light baryons. 
They essentially outline how longitudinal momentum is distributed among the partons within a light baryon's leading Fock state. Alongside parton distribution functions (PDFs), which detail the parton distribution within baryons, they jointly provide a comprehensive description of baryonic structure.
Moreover, light baryon LCDAs also play a important role in both standard model investigations \cite{LHCb:2015eia} and explorations of new physics \cite{FermilabLattice:2015cdh,LHCb:2015tgy,LHCb:2021byf}.
%Light-cone distribution amplitudes (LCDAs) of light baryons have always been an important building block for describing light baryons.
%By their definition one can see that they are important for understanding the internal structure of the light baryons.
%The LCDAs which characterize the assignment of the longitudinal momentum among the partons within the leading Fock state of the light baryons, and the PDF(parton distribution function) which describes the distribution of the parton in baryons, give a description for a baryon structure together.
%In addition, light baryon LCDAs also play an important role in the study of standard model [] and new physics [].

Despite their significance, light baryon LCDAs have not gained as much attention as PDFs. 
The primary challenge stems from the fact that 
in an exclusive process, several LCDAs through the convolution integrals enter the same physical observable.
Additionally, for light baryons, their LCDAs are dependent on two variables, setting them apart from the more straightforward investigations of PDFs and meson DAs cases.

Historically, researches on light baryon LCDAs have predominantly relied on QCD sum rules \cite{Chernyak:1984bm,King:1986wi,Braun:1999te,Anikin:2013aka} and lattice QCD \cite{Gockeler:2008xv,QCDSF:2008qtn,Bali:2015ykx,RQCD:2019hps}. 
Given their inherent non-perturbative nature, the results are model-dependent and entail uncontrollable uncertainties. 
Consequently, only the lowest moments have been obtained \cite{QCDSF:2008qtn,Bali:2015ykx,RQCD:2019hps}.

%Despite its importance, thee light baryon LCDAs had not draw as much as attention as PDF.
%The main difficulty lies in the fact that .
%Besides, for the light baryons, their LCDAs depend on two variables, unlike the PDF or meson case, which makes the investigate more complicated.
%In the past, the light baryon LCDAs mainly rely on QCD sum rules[] and Lattice QCD[]. 
%Due to their intrinsic non-perturbative nature, the results are model-dependent and have uncontrollable uncertainties.
%Till then, only the lowest moments can be given[].
In resent papers \cite{Deng:2023csv,Han:2023xbl}, large momentum effective theory (LaMET)  was adopted to study light baryon LCDA from the first principle through lattice QCD.
LaMET has been employed to investigate various quantities, including
PDFs \cite{Xiong:2013bka,Chen:2016utp,Xiong:2017jtn,Wang:2017qyg,Chen:2017mzz,Liu:2018hxv,Fan:2018dxu,Ebert:2018gzl,Zhang:2018diq,LatticeParton:2018gjr,Chen:2018xof,Zhang:2018nsy,Chai:2019rer,Cichy:2019ebf,Izubuchi:2019lyk,Bhattacharya:2020jfj,Bhattacharya:2020cen,Chai:2020nxw,Shugert:2020tgq,Lin:2020ssv,Alexandrou:2020qtt,Bhattacharya:2021moj,Bhattacharya:2021rua,LatticeParton:2022xsd,Zhang:2023bxs},
meson LCDAs
\cite{Zhang:2017bzy,Bali:2018spj,Xu:2018mpf,Zhang:2017zfe,Liu:2018tox,Wang:2019msf,Zhang:2020gaj,Hua:2020gnw,hua_pion_2022,Hu:2023bba},
TMDs
\cite{Ji:2014hxa,Ji:2018hvs,Ji:2019ewn,Ji:2019sxk,Ebert:2019okf,ebert_one-loop_2020,ji_transverse-momentum-dependent_2020,shanahan_lattice_2021,LatticeParton:2020uhz,Li:2021wvl,schindler_one-loop_2022,zhang_renormalization_2022,zhu_gluon_2022,LPC:2022zci,LatticePartonLPC:2023pdv,Zhao:2023ptv},
GPDs
\cite{Ji:2015qla,Liu:2019urm,Chen:2019lcm,alexandrou_unpolarized_2020,lin_nucleon_2021,alexandrou_generalized_2021,Dodson:2021rdq,alexandrou_transversity_2022,scapellato_proton_2022,Ma:2022gty,Bhattacharya:2023nmv},
LFWFs
\cite{ji_computing_2022},
TMDWFs
\cite{Deng:2022gzi,Chu:2022uyk,LPC:2022ibr,Chu:2023jia},
and DPDs
\cite{Zhang:2023wea,jaarsma_towards_2023}, demonstrating its capability in the study of light-cone quantities.
For more papers on the content and applications of LaMET, please refer to \cite{Cichy:2018mum,Zhao:2018fyu,Ji:2020ect} and the references therein.
In paper \cite{Deng:2023csv}, the calculation was performed in momentum space, and the corresponding quasi-distribution amplitude (quasi-DA) was renormalized using the RI/MOM scheme. 
However, despite it is adoptable theoretically, the application of this scheme on the lattice introduced uncontrollable infrared effects.
%, rendering the resulting kernel ill-defined.
To address theses issues, we have developed a hybrid renormalization scheme specifically designed to handle the spatial correlator's divergences in different coordinate regions \cite{Han:2023xbl}. 

However, throughout the entire process, the spatial correlator has not been comprehensively introduced with regard to its structures. 
To address this gap, this paper focuses on a detailed one-loop analysis of the spatial correlators of light baryons.
In conjunction with the introduction of the calculation processes, this paper will also introduce the relevant ultraviolet (UV) and infrared (IR) structures inherent in the correlators.
In this paper, the spatial correlators will be renormalized using the ratio scheme, which involves renormalizing spatial correlations by dividing by their own 0-momentum matrix element \cite{Orginos:2017kos,Radyushkin:2017cyf,Radyushkin:2017lvu}. 
%The ratio scheme is primarily designed for short-distance applications.
Besides, by employing the Ioffe-time distribution definition, the light-cone correlator and spatial correlator can be studied on an equal footing.
After that, we will perform the matching between the spatial correlator and light-cone correlator directly on the coordinate space.
Then the LCDAs can be obtained by performing the Fourier transformation upon the light-cone correlator.
The Renormalization Group Equation (RGE) for the LCDA is also provided, and their connections with spatial correlator are discussed.

The rest of the paper is arranged as follows.
Section.~\ref{sec-set} covers the essential content related to LCDAs and spatial correlators.
Section.~\ref{sec-oneloop} is dedicated to the calculation of one-loop results, where we present the patterns involved in spatial correlation calculations and analyze their UV and IR structures.
Section.~\ref{sec-ratio} focuses on the renormalization process through the ratio scheme, followed by matching. 
Additionally, we provide the scaling behavior of the LCDA for comparison with previous results as a validation check.
%The renormalization through ratio scheme and the following matching are performed in the section 4.
%The scaling behavior of the LCDA has also been given for comparing with before results as a check.
The paper is summarised in the last section.

%The paper concludes with a summary in the last section.

\section{Light-cone distribution amplitudes and Spatial correlators for a light baryon}
\label{sec-set}

In this section, we introduce the requisite notations and conventions required for subsequent discussions.  
In particular, the definition of LCDAs and Ioffe-time distribution (ITD) will be given.
We start with the LCDAs, which are defined as the hadron-to-vacuum matrix elements of non-local operators consisting of quarks and gluon which live on the light cone.
In the case of a light baryon, the three-quark matrix element can be constructed as~\cite{Braun:1999te}
\begin{widetext}
 \begin{equation}
\left\langle 0\left|\varepsilon^{i j k} u_\alpha^{i^{\prime}}\left(z_1 \right)U_{i^{\prime} i}\left(z_1 , z_0 \right) d_\beta^{j^{\prime}}\left(z_2 \right)
U_{j^{\prime} j}\left(z_2 , z_0 \right) 
s_\gamma^{k^{\prime}}\left(z_3 \right)
U_{k^{\prime} k}\left(z_3 , z_0 \right)
\right| \Lambda(P, \lambda)\right\rangle,
\end{equation}   
\end{widetext}
where $\left.\left.\right | \Lambda(P, \lambda)\right\rangle$ stands for the $\Lambda$ baryon state with the momentum $P$,  $P^2=0$ and the helicity $\lambda$.
$\alpha$, $\beta$ and $\gamma$ are Dirac indices.
$i^{(\prime)}$, $j^{(\prime)}$ and $k^{(\prime)}$ denote color charges.
%$z_i$s are set on the light cone, $z_i^2=0$.
In this paper, two light-cone unit vectors are defined as $n^\mu=(1,0,0,-1)/\sqrt{2}$ and $\bar n^\mu=(1,0,0,1)/\sqrt{2}$. 
The momentum of the baryon is along the $\bar n$ direction, $P^\mu=P^{+} \bar n^\mu = (P^z,0,0,P^z)$.
The coordinates are set in the $n$ direction, $z_i^\mu=z_i n^\mu$.  
The Wilson lines $U(x,y)$
\begin{equation}
U(x, y)=\mathcal{P} \exp \left[i g \int_0^1 \mathrm{~d} t(x-y)_\mu A^\mu(t x+(1-t) y)\right]
\end{equation}
are inserted to preserve the gauge invariance. 
For simplicity and brevity, we will choose $z_0=0$.
Besides, the Wilson lines, color indexes, and helicity will not be written out explicitly below.

Based on Lorentz invariance,  and the spin and parity requirement, the matrix element can be decomposed in terms of three functions, $V(z_i P\cdot n)$, $A(z_i P\cdot n)$, and $T(z_i P\cdot n)$ to the leading twist
\begin{align}
& \left\langle 0\left|u_\alpha^{}\left(z_1\right) d_\beta^{}\left(z_2\right) s_\gamma^{}\left(z_3\right) \right| \Lambda(P)\right\rangle 
\\
& ={f_N}\left\{( P\!\!\!\!/ C)_{\alpha \beta}\left(\gamma_5 u_{\Lambda}\right)_\gamma V\left(z_i P\cdot n\right) \right.
\\&
+\left( P\!\!\!\!/ \gamma_5 C\right)_{\alpha \beta} (u_{\Lambda})_\gamma A\left(z_i P\cdot n\right)
\\&
\left.+\left(i \sigma_{\mu \nu} P^\nu C\right)_{\alpha \beta}\left(\gamma_\mu \gamma_5 u_{\Lambda}\right)_\gamma T\left(z_i P\cdot n\right)\right\} ,\notag
\end{align}
where $C$ signifies the charge conjugation.
$u_{\Lambda}$ stands for the $\Lambda$ baryon spinor. Equivalently,  the three leading twist functions can be projected by inserting a specific gamma matrix $\Gamma$ into the $u$ and $d$ quark fields. 
In the following discussion, we will take $A(z_i P \cdot n)$ as an example while the other matrix elements can be similarly analyzed. 
Then we have 
\begin{equation}\label{eq:LCDA}
\begin{aligned}
& \widetilde{\mathcal I} (z_1,z_2,z_3,P^+,\mu)=
\left\langle
    0\left|
        \psi_1^T\left(z_1\right) {\Gamma} \psi_2\left(z_2\right) \psi_3\left(z_3\right)
    \right| \Lambda(P)
\right\rangle_R,
\\&
{\Phi_L}\left(x_1, x_2, \mu \right) f_{\Lambda}(\mu) P^{+} u_{\Lambda}(P)
\\&
=\int_{-\infty}^{+\infty} \frac{d \, P^+ z_1}{2 \pi} \frac{d \, P^+ z_2}{2 \pi} e^{i x_1 P^+ z_1+i x_2 P^+ z_2} \widetilde{\mathcal I}(z_1,z_2,0,P^+,\mu),
\end{aligned}
\end{equation}
where $T$ means transpose and ${\Gamma}=C \gamma_5 \slashed n $.  
$R$ stands for renormalization. $x_i$s label the longitudinal momentum fractions carried by the three quarks and $0\leq x_i \leq 1$. The $\mu$ denotes the renormalization scale which will be converted to the factorization scale when the factorization of quasi-DA is established. $f_\Lambda(\mu)$ is the $\Lambda$ baryon decay constant defined as follows
$f_\Lambda(\mu) P^{+} u_{\Lambda}(P) = \widetilde{\mathcal I}(0,0,0,P^+,\mu).$
%It should be noted that we have defined the LCDA ${\Phi_L}\left(x_1, x_2, \mu \right)$ by separating the baryon decay constant $f_\Lambda(\mu)$, which has a different convention with the recent LQCD calculation~\cite{RQCD:2019hps}. 
Note that $f_\Lambda(\mu)$ depends on the renormalization scale $\mu$ since the local operator here is not a conserved current.   The LCDA ${\Phi_L}\left(x_1, x_2, \mu \right)$ in Eq.~(\ref{eq:LCDA}) is dimensionless and normalized.   
%It should be mentioned that the LCDAs can be expanded in the orthogonal polynomials basics~\cite{Anikin:2013aka}
%\begin{equation}
%\varphi_N\left(x_i, \mu\right)=120 x_1 x_2 x_3 \sum_{n=0}^{\infty} \sum_{k=0}^n \varphi_{n k}^N(\mu) \mathcal{P}_{n k}\left(x_i\right).
%\end{equation}

For the lattice QCD side, in order to extract the LCDA, 
the first step involves selecting an appropriate spatial correlator. 
In this paper, the spatial correlator is chosen as \cite{Braun:1999te}
\begin{equation} \label{eq:MDA}
   % \begin{aligned}
        \widetilde M(z_1,z_2,z_3,P^{z},\mu)=\left\langle 0\left|u^T\left(z_1\right) \widetilde{\Gamma} d\left(z_2\right) s\left(z_3\right)\right| \Lambda(P)\right\rangle_R,
   % \end{aligned}
\end{equation}
where $\tilde \Gamma= C \gamma_5 n\!\!\!\slash_z$.
And the coordinates are set as $z_i^\mu=z_i n_z^\mu $, where $n_z^\mu=(0,0,0,1)$.

These two kinds of corrlators can be treated in a more unified manner.
The light-cone correlator can be understood as a function of two Lorentz-invariant arguments, $z_i P \cdot n$ and $z^2$.
It has been extended to distributions beyond those lying on the light cone and is referred to as Ioffe-time distribution (ITD) \cite{Braun:1994jq,Radyushkin:2017lvu}.
The ITD is dependent on two Lorentz scalars, Ioffe time $\nu_i$ (defined as $-z_i P\cdot n_i$) and distance $z_i^2$, where the specific values of $n_i$ rely on our requirements.
Consequently, we can represent the light-cone correlator and spatial correlator as
\begin{equation}
\begin{aligned}
    \mathfrak{I}(z_1,z_2,z_3,P^+,\mu)&\equiv\mathfrak{I}(\nu_1,\nu_2,\nu_3,z^2,\mu),
    \\
    \mathfrak{M}(z_1,z_2,z_3,P^z,\mu)&\equiv\mathfrak{M}(\nu_1,\nu_2,\nu_3,z^2,\mu),
    \end{aligned}
\end{equation}
with $z^2$ being an expression compactly representing all possible contractions of the $z_i$ terms.
It should be noted that, for the spatial correlator, only the leading twist component will be retained,  which means that we will only keep the part that is proportional to $P^\mu$.

To establish a connection between the LCDAs and spatial correlators, in contrast to previous approaches, we directly extract the LC correlator by matching it with the spatial correlator in coordinate space. 
Subsequently, the light-cone correlator can be Fourier-transformed into LCDAs.

\section{One-loop calculation of the spatial correlator in coordinate space }\label{sec-oneloop}
\begin{figure}[htb]
\centering
\includegraphics[width=0.30\textwidth]{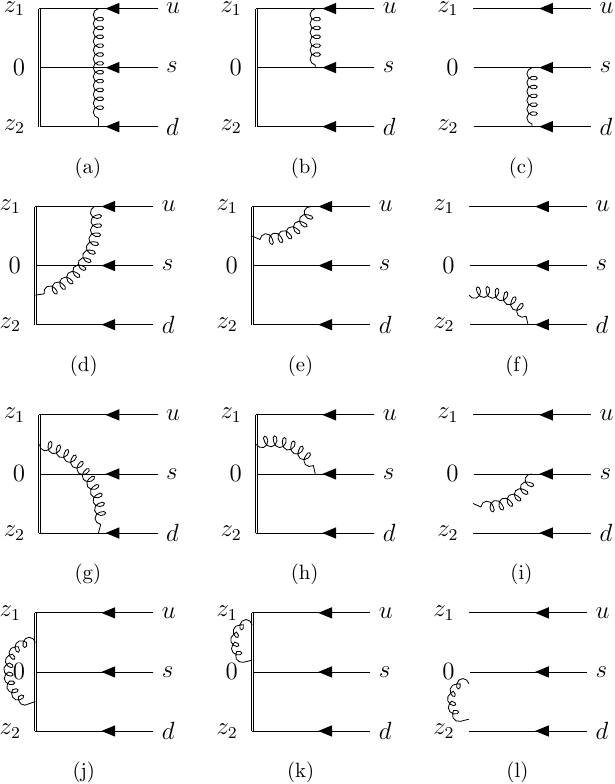}
\caption{One loop corrections for the equal-time matrix element of the $\Lambda$ baryon.}
\label{Pspic}
\end{figure}

%To obtain the quasi-DAs from lattice simulations, it is necessary to separately consider their properties in the short and long-distance regimes. 
%From this perspective, working in coordinate space offers advantages.
%where the lower index ``$p$" in $M_{p}$ denotes the perturbative calculation. 
In this section, the one-loop results for the spatial correlator and the light-cone correlator will be presented.
These results will be presented in dimensional regularization with $\overline{\text{MS}}$ renormalization. 
%The results are gauge invariant, and the Feynman gauge will be adopted in the practical calculation.
We will stick to adopt the Feynman gauge throughout, though the results are gauge invariant.
All the calculation will be performed on the operator level.
By following this approach, the desired matrix elements can be obtained by incorporating suitable out-states.
%For instance, we will take the leading Fock state of the $\Lambda$ baryon as an example by sandwiching the operators between the vacuum state $\left \langle0|\right.$ and the lowest-order Fock state $\left.|uds\right \rangle$:
%\begin{align}
%&{M}(z_1,z_2,z_3=0,P^z,\mu) 
%\\&
%= \left\langle 0\left|\psi_1^T\left(z_1\right) \widetilde\Gamma \psi_2\left(z_2\right) \psi_3\left(0\right)\right| u(x_1 P)d(x_2 P)s(x_3 P)\right\rangle_R.    
%\end{align}
As shown in Fig.~\ref{Pspic}, there are twelve distinct diagrams to calculate, which can be divided into three categories: quark-quark (q-q), quark-Wilson line (q-W), and Wilson line-Wilson line (W-W).
We will begin with the q-W pattern, which exhibits the most complex structures among all three patterns. 
In the q-W pattern, there are two different situations to consider: cases (e), (f), (h), and (i), and cases (d) and (g).
%%%%%%%%%%%%%%%%%%%%%%
%\begin{widetext}
%\end{widetext}

%%%%%%%%%%%%%%%%%%%%%%
\begin{widetext}
\subsection{q-W pattern}
\subsubsection{\textbf{Cases~(e), (f), (h), and (i)}}

We take Fig.~\ref{Pspic}(e)  as the example to illustrate the calculation, in which the one-loop corrections are
 \begin{equation} 
 \widetilde{O}_e=
 \left(\psi _1 \left(z_1\right) \left(i g_s \int d^d \eta _1 \bar \psi _1 \left(\eta _1\right) \slashed A \left(\eta _1\right) \psi _1 \left(\eta _1\right)\right)\right){}^T 
 \left(-i g_s \int_0^1 d t_1 z_1 \cdot A \left(t_1 z_1\right)\right) \widetilde \Gamma  \psi _2 \left(z_2\right) \psi _3 (0) . 
 \end{equation} 
     
%\end{widetext}
The color indexes and the parameter $\left(\displaystyle\frac{\mu ^2}{e^{\ln (4 \pi )-\gamma_E}}\right)^{\epsilon }$ are not written out explicitly.
The gluon and quark propagators in  the coordinate space are 
\begin{eqnarray} 
 G(x-y)=\frac{\Gamma(d / 2-1)}{4 \pi^{d / 2}} \frac{-g_{\mu \nu}}{\left(-(x-y)^2+i \epsilon\right)^{d / 2-1}}; \quad\quad
Q(x-y)=\frac{\Gamma(d / 2)}{2 \pi^{d / 2}} \frac{i(\slashed x-\slashed y)}{\left(-(x-y)^2+i \epsilon\right)^{d / 2}}. 
\end{eqnarray}

%\begin{widetext}
Following the standard routine, substituting them back and rearranging the formula, one arrives at
\begin{align}
 \widetilde{O}_e&= g_s^2 \frac{(-i)^{d / 2-1}}{8 \pi^{d / 2}} \int d^d k_1 \int_0^1 d t_1 \int_0^\infty d \sigma_1 \int_0^\infty  d\sigma_2 \sigma_1^{d / 2-1} \sigma_2^{d / 2-2}\left(\sigma_1+\sigma_2\right)^{-d / 2} 
 \\
 &\times  e^{\frac{i\left(4 \left(\sigma_1+\sigma_2 t_1\right)z_1(k_1 \cdot n_z)+k_1^2-4 \sigma_1 \sigma_2\left(t_1-1\right)^2 (-z_1^2))\right)}{4\left(\sigma_1+\sigma_2\right)}}  \psi_1^T\left(k_1\right) \left((-z_1^2)-\frac{z_1(k_1 \cdot n_z)+\left(\sigma_1+\sigma_2 t_1\right)(-z_1^2)}{\sigma_1+\sigma_2}\right) \widetilde \Gamma\psi_2\left(z_2\right) \psi_3(0),\notag
\end{align}    
\end{widetext}
where $\sigma_1$ and $\sigma_2$ are Schwinger parameters, and $k_1$ is from the Fourier transformation of $\psi_1^T(z_1)$.
Note that terms like  $k_1^2$ or $\slashed k_1 \psi (z)$ have been neglected in the calculation due to the equation of motion. 
By changing $(\sigma_1,\sigma_2)$ to $(\sigma,\eta)$ with $\sigma_1=\displaystyle\frac{\sigma}{\eta}$ and $\sigma_2=\displaystyle\frac{\sigma}{1-\eta}$, the above result can be rearranged as
\begin{widetext}
    
\begin{align}
\widetilde{O}_e=&\widetilde{O}_{e1}+\widetilde{O}_{e2}
\\
 \widetilde{O}_{e1}=
 &-g_s^2 \frac{1}{8 \pi^{d / 2}} \Gamma(d / 2-1) \int d^d k_1 \int_0^1 d t_1 \int_0^1 d \eta \left(1-t_1\right)^{3-d}\left(z_1^2\right)^{2-d / 2} 
 \\&
 \times e^{i z_1 k_1\left(\eta\left(t_1-1\right)+1\right)} \psi_1^T\left(k_1\right) \widetilde\Gamma  \psi_2\left(z_2\right) \psi_3(0),
\\
 \widetilde{O}_{e2}=&-g_s^2 \frac{(-i)^{d / 2-1}}{8 \pi^{d / 2}} \int d^d k_1 \int_0^1 d t_1 \int_0^1 d \eta \int_0^\infty d\sigma \sigma^{\frac{d}{2}-3}\left(1-\eta\right) z_1(k_1 \cdot n_z) 
  \\&
\times e^{-i z_1 n_z \cdot \left(k_1\left(-\left(\eta\left(t_1-1\right)\right)-1\right)+\sigma\left(t_1-1\right)^2 z_1 n_z\right)} 
\psi_1^T\left(k_1\right)\widetilde\Gamma \psi_2\left(z_2\right) \psi_3(0),
\end{align} 

and then one can separate it into two parts and calculate them respectively.

For $\widetilde{O}_{e1}$, we further define
and we can have the simplified form
   
\begin{equation}
\begin{aligned}
 \widetilde{O}_{e1}=-g_s^2 \frac{1}{8 \pi^{d / 2}} \Gamma(d / 2-1)\left(z_1^2\right)^{2-d / 2} &\int_0^1 d t_0 \int_0^1 d \eta\left(t_0\right)^{3-d} 
   \psi_1^T\left(\left(1-\eta t_0\right) z_1\right) \widetilde\Gamma \psi_2\left(z_2\right) \psi_3(0),
\end{aligned}
\end{equation}

\end{widetext}
with $t_0=1-t_1$. 
The $t_0\to 0$ corresponds to a UV divergence since that divergence is regularized by $d<4$ and one end of the Wilson line approaches $z_1$ when $t_0\to 0$.
One can separate this divergence from the rest  by using $\psi^T((1-\eta t_0)z_1)=\left(\psi^T((1-\eta t_0) z_1)-\psi^T(z_1)\right )+\psi^T(z_1)$.
Then it is straightforward to obtain the results for  these two parts  
\begin{widetext}
    
 \begin{eqnarray} 
 \widetilde{O}_{e11}=& \displaystyle\frac{\alpha _s C_F}{4 \pi } \left(\frac{1}{\epsilon _{\text{UV}}}+\log \left(\frac{1}{4} \mu ^2 z_1^2 e^{2 \gamma_E }\right)\right) 
 \psi _1^T \left(z_1\right) \widetilde\Gamma  \psi _2 \left(z_2\right) \psi _3 (0), 
 \\
 \widetilde{O}_{e12}=&\displaystyle \frac{\alpha _s C_F}{2 \pi } \int_0^1 d \eta  \left(\frac{1-\eta }{\eta }\right)  _+ 
 (\psi _1^T \left((1-\eta ) z_1\right) \widetilde\Gamma  \psi _2 \left(z_2\right) \psi _3 (0). 
 \end{eqnarray} 
The plus function is defined as 
$\displaystyle\int_0^1 d u \left[ G(u) \right]_{+} F(u)=\displaystyle\int_0^1 d u G(u) [F(u)-F(0)].$
 
%\end{widetext}

%\paragraph{\textbf{Term involving $z_1(k_1 \cdot n_z)$.}}
For $\widetilde{O}_{e2}$,
%\begin{widetext}
there is an IR divergence: 
 \begin{eqnarray}
 \begin{aligned}
     \widetilde{O}_{e2}=&-C_F \frac{g_s^2}{8 \pi^2} \int_0^1 d \eta\left(\left(\ln \left(\frac{1}{4} \mu^2 z_1^2 e^{2 \gamma_E}\right)+\frac{1}{\epsilon_{\mathrm{IR}}}+2\right)\left(\frac{1-\eta}{\eta}\right)_{+}+\left(\frac{2 \ln \eta} {\eta}\right)_{+}\right) 
%\\&\times
\psi_1^T\left(z_1(1-\eta)\right) \widetilde\Gamma  \psi_2\left(z_2\right) \psi_3(0).
 \end{aligned}
\end{eqnarray}    

%\end{widetext}
Collecting all these pieces and removing the UV divergence in the $\overline{\rm MS}$ scheme give the final result: 
%\clearpage
%\begin{widetext}
 \begin{equation} 
 \begin{aligned} 
 \widetilde{O}_e&=
 \frac{\alpha_s C_F}{4 \pi} \ln \left(\frac{1}{4}\mu_{\text{UV}} ^2 z_1^2 e^{2 \gamma_E}\right)  \psi _1^T \left(z_1\right) \Gamma  \psi _2 \left(z_2\right) \psi _3 (0)
 -\frac{\alpha_s C_F}{ \pi} \int_0^1 d \eta \left(\frac{\ln \eta }{\eta } \right) _+ \psi _1^T \left((1-\eta ) z_1\right) \widetilde\Gamma  \psi _2 \left(z_2\right) \psi _3 (0)
 \\& 
-\frac{\alpha_s C_F }{2 \pi} \int_0^1 d \eta \left(\frac{1-\eta }{\eta } \right) _+ \left(\ln \left(\frac{1}{4}\mu_{\text{IR}} ^2 z_1^2 e^{2 \gamma_E }\right)+\frac{1}{\epsilon _{\text{IR}}}+1\right) \psi _1^T \left((1-\eta ) z_1\right) \widetilde\Gamma  \psi _2 \left(z_2\right) \psi _3 (0) ,
 %\\&  
 \end{aligned} 
 \end{equation}   
%\end{widetext}
where $\displaystyle\alpha_s=\frac{g^2_s}{4 \pi}$.
All the other cases will renormalized in this manner below without mention.
%We have checked that after making a Fourier transformation, the above results are consistent with Ref.~\cite{Deng:2023csv} in momentum space. 
Then, in the same manner, results for the  quark-Wilson-line diagrams are derived as: 
cases~(e), (f), (h), and (i) all can be derived:
%\begin{widetext}

\begin{equation}
    \begin{aligned}
    \widetilde{O}_{e}&= 
    \frac{\alpha_s C_F}{4 \pi} 
    %\ln \left(\frac{1}{4}\mu^2 z_1^2 e^{2 \gamma_E}\right)
    L_1^{\text{UV}}
    \psi_1^{T}\left(z_1\right) \widetilde\Gamma \psi_2\left(z_2\right) \psi_3(0) 
    \\
    & -\frac{\alpha_s C_F}{2 \pi} \int_0^1 d \eta \psi_1^{T}\left((1-\eta) z_1\right) \widetilde\Gamma \psi_2\left(z_2\right) \psi_3(0) 
    %\\& \times 
    \left\{
    %\ln \left(\frac{1}{4}\mu^2_{\text{IR}} z_1^2 e^{2 \gamma_E+1}\right)
    \left(L_1^{\text{IR}}+1+\frac{1}{\epsilon_{\mathrm{IR}}}\right)
    \left(\frac{1-\eta}{\eta}\right)_{+}
    +2\left(\frac{\ln \eta}{\eta}\right)_{+}\right\},
%\end{aligned}
    %\end{equation} 
%\begin{equation}
    %\begin{aligned}
    \\
    \widetilde{O}_{h}&= 
     \frac{\alpha_s C_F}{8 \pi} 
    %\ln \left(\frac{1}{4}\mu^2 z_1^2 e^{2 \gamma_E}\right)
    L_1^{\text{UV}}
    \psi_1^{T}\left(z_1\right) \widetilde\Gamma \psi_2\left(z_2\right) \psi_3(0) 
    \\
    & -\frac{\alpha_s C_F}{4 \pi} \int_0^1 d \eta \psi_1^{T}\left(z_1\right) \widetilde\Gamma \psi_2\left(z_2\right) \psi_3\left(\eta z_1\right) 
    %\\& \times
    \left\{
    %\left(\ln \left(\frac{1}{4}\mu^2_{\text{IR}} z_1^2 e^{2 \gamma_E+1}\right)\right)
    \left(L_1^{\text{IR}}+1+\frac{1}{\epsilon_{\mathrm{IR}}}\right)
    \left(\frac{1-\eta}{\eta}\right)_+
    +2\left(\frac{\ln \eta}{\eta}\right)_{+}\right\},
    \\
        \widetilde{O}_{f}&= 
         \frac{\alpha_s C_F}{4 \pi} 
        L_2^{\text{UV}}
        \psi_1^{T}\left(z_1\right) \widetilde\Gamma \psi_2\left(z_2\right) \psi_3(0) 
        \\
        & -\frac{\alpha_s C_F}{2 \pi} \int_0^1 d \eta \psi_1^{T}\left( z_1\right) \widetilde\Gamma \psi_2\left((1-\eta) z_2\right) \psi_3(0) 
        %\\& \times
        \left\{
        \left(L_2^{\text{IR}}+1+\frac{1}{\epsilon_{\mathrm{IR}}}\right)
        \left(\frac{1-\eta}{\eta}\right)_{+}
        +2\left(\frac{\ln \eta}{\eta}\right)_{+}\right\} ,
\\
        \widetilde{O}_{i}&= 
         \frac{\alpha_s C_F}{8 \pi} 
        L_2^{\text{UV}}
        \psi_1^{T}\left(z_1\right) \widetilde\Gamma \psi_2\left(z_2\right) \psi_3(0) 
        \\
        & -\frac{\alpha_s C_F}{4 \pi} \int_0^1 d \eta \psi_1^{T}\left(z_1\right) \widetilde\Gamma \psi_2\left(z_2\right) \psi_3\left(\eta z_2\right) 
        \left\{
        \left(L_2^{\text{IR}}+1+\frac{1}{\epsilon_{\mathrm{IR}}}\right)
        \left(\frac{1-\eta}{\eta}\right)_+
        +2\left(\frac{\ln \eta}{\eta}\right)_{+}\right\}.
    \end{aligned}
\end{equation}  
%\end{widetext}
Some abbreviations are used in the above:
\begin{eqnarray}
L_1^{\text{IR, UV}}=\ln \left(\displaystyle\frac{1}{4}\mu_{\text{IR, UV}} ^2 z_1^2 e^{2 \gamma_E }\right); %\;\;\;
L_2^{\text{IR,UV}}=\ln \left(\displaystyle\frac{1}{4}\mu_{\text{IR,UV}} ^2 z_2^2 e^{2 \gamma_E }\right); 
%\\
L_{12}^{\text{IR,UV}}=\ln\left(\displaystyle\frac{1}{4}\mu^2_{\text{IR,UV}}(z_1-z_2)^2 e^{2\gamma_E}\right). 
\end{eqnarray}
Since the color parameters for any chosen baryon out-states are fixed, 
we have preincluded these color parameters in the operator expressions to simplify the formulas.

\subsubsection{\textbf{Cases~(d) and (g)}}
%\begin{widetext}
There are more subtleties in cases~(d) and (g).
We take case~(d) 
%\begin{widetext}
\begin{equation}
\widetilde{O}_d=
\psi_1(z_1) (ig\mu^{\frac{4-d}{2}}\int d^d\eta_1 \bar{\psi}_1(\eta_1)A\!\!/(\eta_1)\psi_1(\eta_1))\widetilde\Gamma (-ig\mu^{\frac{4-d}{2}}\int_0^1 dt_1 z_2\cdot A(t_1z_2)) \psi_2(z_2)\psi_3(0),
\end{equation}
as a demonstration to illustrate them.
%\end{widetext}
Substituting and arranging as in the previous cases, the case~(d) can be separated into two parts: 
%\begin{widetext}
\begin{align}
\widetilde{O}_d=&\widetilde{O}_{d1}+\widetilde{O}_{d2},
\\
    \widetilde{O}_{d1}=&g^2\frac{(-i)^{d/2-1}}{8\pi^{d/2}}\mu^{4-d}
    \int d^d k_1\int_0^1 dt_1 \int_0^1 d \beta_1\int_0^\infty d\sigma   \sigma ^{\frac{d}{2}-2}
    (-\left(z_1-t_1 z_2\right)z_2)
\\&
    \times e^{-i \left(k_1 \left(\left(\beta _1-1\right) z_1-\beta _1 t_1
    z_2\right)+\sigma  \left(z_1-t_1 z_2\right){}^2\right)}\psi_1^T(k_1)Cn\!\!/\gamma_5\psi_2(z_2)\psi_3(0),
\\
    \widetilde{O}_{d2}=&g^2\frac{(-i)^{d/2-1}}{8\pi^{d/2}}\mu^{4-d}
    \int d^d k_1\int_0^1 dt_1\int_0^1 d \beta_1\int_0^\infty d\sigma   \sigma ^{\frac{d}{2}-3} \left(\beta _1-1\right) 
    (k_1\cdot z_2)
\\&
    \times e^{-i \left(k_1 \left(\left(\beta _1-1\right) z_1-\beta _1 t_1 z_2\right)+\sigma  \left(z_1-t_1 z_2\right)^2\right)}\psi_1^T(k_1)Cn\!\!/\gamma_5\psi_2(z_2)\psi_3(0).
\end{align}
%\end{widetext}
Through the calculation in the previous case, we now know that the IR and UV divergence have been separated during this operation.

For the $\widetilde {O}_{d1}$,
%involving $\sigma(z_1-t_1 z_2)z_2$, 
to make the inside two forms more explicitly, 
the range of the integral is split further as
%\clearpage
%\begin{widetext}
\begin{equation}
\begin{aligned}
\widetilde{O}_{d1}=&\widetilde{O}_{d11}-\widetilde{O}_{d12}
\\
=&g^2\frac{\Gamma(d/2-1)}{8\pi^{d/2}}\mu^{4-d}(\int_{\frac{z_1}{z_2}}^1 dt_1-\int_{\frac{z_1}{z_2}}^0 dt_1)
\int_0^1 d\beta_1
   (-\left(z_1-t_1 z_2\right)z_2)((z_1-t_1z_2)^2)^{1-d/2}
\\&
\times\psi_1^T((1-\beta _1) z_1+\beta _1 t_1
   z_2)Cn\!\!/\gamma_5\psi_2(z_2)\psi_3(0).
\end{aligned}
\end{equation}    
%\end{widetext}

Following that, we redefine the integral variables within them respectively, 
%after rearrangement, we get
%\begin{widetext}
 \begin{equation}
\begin{aligned}
\widetilde{O}_{d11}&=
g^2\frac{\Gamma(d/2-1)}{8\pi^{d/2}}\frac{((z_1-z_2)^2)^{2-d/2}}{d-3}\mu^{4-d}
\int_0^1d\eta(\eta^{3-d}-1)
%\\&
\psi_1^T((1-\eta _1) z_1+\eta _1
   z_2)Cn\!\!/\gamma_5\psi_2(z_2)\psi_3(0),
\\
\widetilde{O}_{d12}&=
g^2\frac{(z_1^2)^{2-d/2}}{d-3}\frac{\Gamma(d/2-1)}{8\pi^{d/2}}\mu^{4-d}
\int_0^1d\eta (\eta^{3-d}-1)
%\\&
\psi_1^T((1-\eta_1) z_1)(-Cn\!\!/\gamma_5)\psi_2(z_2)\psi_3(0).
\end{aligned}
\end{equation}
%\end{widetext}
Now one can see that the two distinct forms of contributions have be separated.

For the $\widetilde {O}_{d2}$ ,
%involving $k_1 z_2$,
%\begin{widetext}
%\end{widetext}
it can be divided into two parts: 
%one is depends on $z^2$ and another one depends on $z \cdot k$,
%\begin{widetext}
\begin{equation}
\begin{aligned}
&\widetilde O_{d2}=\widetilde O_{d21}-\widetilde O_{d22},
\\&
\widetilde O_{d21}=
\Gamma(d/2-2)g^2\frac{-1}{8\pi^{d/2}}\mu^{4-d}\int_0^1d\beta_1\int_0^1 dt_1z_2(z_1-t_1z_2)^{-d+3}\psi_1^T((1-\beta_1)z_1+t_1\beta_1z_2)Cn\!\!/\gamma_5\psi_2(z_2)\psi_3(0),
\\&
\widetilde O_{d22}=-\Gamma(d/2-2)g^2\frac{-1}{8\pi^{d/2}}\mu^{4-d}\int d^d k_1\int_0^1d\beta_1\int_0^1 dt_1z_2(z_1-t_1z_2)^{-d+3}
e^{i(k_1 \cdot z_1)}
\psi_1^T(k_1)Cn\!\!/\gamma_5\psi_2(z_2)\psi_3(0).
\end{aligned}
\end{equation}    
%\end{widetext}

%Once again, we need to split it into two parts.
For $\widetilde O_{d21}$, it can be further divided into two parts by splitting the range of the integral as before:
%\begin{widetext}
 \begin{equation}
\begin{aligned}
\widetilde O_{d21}=&\widetilde O_{d211}-\widetilde O_{d212},
\\
\widetilde O_{d211}=&\Gamma(d/2-2)g^2\mu^{4-d}\frac{-1}{8\pi^{d/2}}
\int_0^1d\eta \int_{\frac{z_1-(z_1-z_2)\eta}{z_2}}^1dt_1
z_2(z_1-t_1z_2)^{-d+3}\frac{z_1-z_2}{z_1-t_1z_2}\\
&
\times
\psi_1^T((1-\eta)z_1+\eta_1z_2)Cn\!\!/\gamma_5\psi_2(z_2)\psi_3(0),
\\
\widetilde O_{d212}=&
\Gamma(d/2-2)g^2\frac{-1}{8\pi^{d/2}}\mu^{4-d}
\int_0^1d\eta\int_{\frac{z_1}{z_2}}^0 dt_1
z_2(z_1-t_1z_2)^{-d+3}%\\
%&
%\times 
\psi_1^T((1-\eta)z_1+t_1\eta z_2)Cn\!\!/\gamma_5\psi_2(z_2)\psi_3(0).
\end{aligned}
\end{equation}   
%\end{widetext}

After redefining the integral variables, above results can be computed as follows:
\begin{equation}
\begin{aligned}
\widetilde O_{d211}=&\Gamma(d/2-2)g^2\frac{-1}{8\pi^{d/2}}\mu^{4-d}((z_1-z_2)^2)^{\frac{4-d}{2}}
\int_0^1d\eta\frac{1-\eta ^{3-d}
   }{d-3}
%\\&
   \psi_1^T((1-\eta)z_1+\eta z_2)Cn\!\!/\gamma_5\psi_2(z_2)\psi_3(0),
\\
\widetilde O_{d212}=&-\Gamma(d/2-2)g^2\frac{-1}{8\pi^{d/2}}\mu^{4-d}(z_1^2)^{\frac{4-d}{2}}\int_0^1d\eta \frac{ \left(1-\eta ^{3-d}\right) }{d-3}
%\\&
\psi_1^T((1-\eta)z_1)(-Cn\!\!/\gamma_5)\psi_2(z_2)\psi_3(0).
\end{aligned}
\end{equation}    

Next, we turn back to consider $\widetilde O_{d22}$, which can be given directly
\begin{equation}
\begin{aligned}
\widetilde O_{d22}=-\mu^{4-d}\Gamma(d/2-2)g^2\frac{-1}{8\pi^{d/2}}\frac{((z_1-z_2)^2)^{2-d/2}-(z_1^2)^{2-d/2}}{d-4}
\psi_1^T(z_1)Cn\!\!/\gamma_5\psi_2(z_2)\psi_3(0).
\end{aligned}
\end{equation}
By summing all these pieces, one can obtain the $\widetilde {O}_{d}$.
Then, for the two cases in this pattern, we have:
\begin{align}
 %%%%%%%%%%%%%%%%%%%% uw2 %%%%%%%%%%%%%%%%%%%%%%%%%%%%%
	 \widetilde{O}_{d}&=
   \frac{\alpha_s C_F}{8}\left(L_{12}^{\text{UV}}-L_{1}^{\text{UV}}\right)\psi_1^T\left( z_1\right) \widetilde\Gamma \psi_2\left(z_2\right) \psi_3(0)
\notag\\
  &-\frac{\alpha_s C_F }{4 \pi}  \int_0^1 d \eta
   \psi_1^T\left((1-\eta) z_1+\eta z_2\right) \widetilde\Gamma \psi_2\left(z_2\right) \psi_3(0)
    %\\&\times 
    \left\{
    \left( L_{12}^{\text{IR}}+1+\frac{1}{\epsilon_{\mathrm{IR}}}\right) 
    \left(\frac{1-\eta}{\eta}\right)_{+}
    +2\left(\frac{\ln \eta}{\eta}\right)_+ \right\}
\notag\\
    &+\frac{\alpha_s C_F }{4 \pi}  \int_0^1 d \eta
    \psi_1^{T}\left((1-\eta) z_1\right) \widetilde\Gamma \psi_2\left(z_2\right) \psi_3(0)
    \left\{
    \left( L_1^{\text{IR}}+1+\frac{1}{\epsilon_{\mathrm{IR}}}\right) 
    \left(\frac{1-\eta}{\eta}\right)_{+}
    +2\left(\frac{\ln \eta}{\eta}\right)_+ \right\}, 
\\
  \widetilde{O}_{g}&=
   \frac{\alpha_s C_F}{8}\left(L_{12}^{\text{UV}}-L_{2}^{\text{UV}}\right)\psi_1^T\left( z_1\right) \widetilde\Gamma \psi_2\left(z_2\right) \psi_3(0)
\notag\\
  &
  -\frac{\alpha_s C_F }{4 \pi} \int_0^1 d \eta
  \psi_1^{T}\left(z_1\right) \widetilde\Gamma \psi_2\left(\eta z_1+(1-\eta) z_2\right) \psi_2\left(z_2\right) \psi_3(0) 
    %\\&\times 
    \left\{
    %\ln \left(\frac{1}{4}
    %\mu_{I R}^2\left(z_1-z_2\right)^2 e^{2 \gamma_E+1} \right)
    \left( L_{12}^{\text{IR}}+1+\frac{1}{\epsilon_{\mathrm{IR}}} \right)
    \left(\frac{1-\eta}{\eta}\right)_{+}
    +2\left(\frac{\ln \eta}{\eta}\right)_+ \right\}
\notag\\
  &+\frac{\alpha_s C_F }{4 \pi} \int_0^1 d \eta
      \psi_1^{T}\left(z_1\right) \widetilde\Gamma \psi_2\left((1-\eta) z_2\right) \psi_3(0)
    %\\&\times 
    \left\{
    %\ln \left(\frac{1}{4}
    %\mu_{I R}^2\left(z_1-z_2\right)^2 e^{2 \gamma_E+1} \right)
    \left( L_2^{\text{IR}}+1+\frac{1}{\epsilon_{\mathrm{IR}}} \right)
    \left(\frac{1-\eta}{\eta}\right)_{+}
    +2\left(\frac{\ln \eta}{\eta}\right)_+ \right\}.
\end{align}    
\end{widetext}
There are no analogous terms in meson cases for these two cases. 
In fact, both of them are combinations of two forms. 
Specifically, there are certain terms in cases~(d) and (g) that have the same form as cases~(e) and (f), respectively. 
It's important to note that there are both infrared (IR) and ultraviolet (UV) singularities in all these cases.
The cases (e) and (f) differ from cases (d), (g), (h), and (i) in terms of their color coefficients, as detailed in \cite{Deng:2023csv}. 
More precisely, for cases (e), (f), (k), and (l), the color algebra yields the same results as in the meson case, represented by $C_F$. 
For cases (d) and (g), the color parameter is $-\displaystyle\frac{C_F}{2}$.

\subsection{q-q pattern}
The quark-quark cases are presented in Fig.~1(a), (b), and (c).
Since there are some differences between case~(a) and cases~(b) and (c), them deserve separate consideration.
\begin{widetext}
\subsubsection{\textbf{Case~(a)}}
In case~(a), after performing the standard procedure, we have
\begin{equation}
\begin{aligned}
 \widetilde{O}_{a}&=\widetilde{O}_{a1}+\widetilde{O}_{a2} ,
\\
    \widetilde{O}_{a1}&=
        (d-2)g^2\mu^{4-d}\frac{(-i)^{d}}{16\pi^{d}}
        \int d^dk_1\int d^dk_2\int d^d\xi_2 \frac{\sigma_1^{d/2-1}\sigma_2^{d/2-1}\sigma_3^{d/2-        2}}{((\sigma_1+\sigma_3))^{d/2}}
        \frac{\sigma _1 \sigma _2 \sigma _3^2
            (-\left(z_1-z_2\right)^2)}{\left(\sigma _2 \sigma _3+\sigma _1\left(\sigma _2+\sigma _3\right)\right){}^2}
\\&
    \times
   e^{\frac{i \left(k_1^2 \sigma _2+\sigma _1 \left(4 \sigma_2 \left(k_1 z_1+k_2 z_2\right)+k_2^2\right)+\sigma _3\left(\left(k_1+k_2\right) \left(k_1+k_2+4 \sigma _1 z_1\right)+4
   \sigma _2 \left(\left(k_1+k_2\right) z_2-\sigma _1
   \left(z_1-z_2\right){}^2\right)\right)\right)}{4 \sigma _2 \sigma
   _3+4 \sigma _1 \left(\sigma _2+\sigma _3\right)}-i \left(\sigma
   _2+\frac{\sigma _1 \sigma _3}{\sigma _1+\sigma _3}\right)
   \xi_2^2}
\\&
    \times 
    \psi_1^T(k_1)Cn\!\!/\gamma^5\psi_2(k_2)\psi_3(0),
\\
%%%%%%%%%%%%%%%%%%%%%%%%%%%%%%%%%%%%%%%%%%
    \widetilde{O}_{a2}&=
    g^2\mu^{4-d}\frac{(-i)^{d}}{16\pi^{d}}
    \int d^dk_1\int d^dk_2\int d^d\xi_2 \int d\sigma_1
    \int d\sigma_2 \int d\sigma_3
    \frac{\sigma_1^{d/2-1}\sigma_2^{d/2-1}\sigma_3^{d/2-1}}{((\sigma_1+\sigma_3))^{d/2+1}}
\\&
    \times 
    e^{\frac{i \left(k_1^2 \sigma _2+\sigma _1 \left(4 \sigma _2 \left(k_1
   z_1+k_2 z_2\right)+k_2^2\right)+\sigma _3
   \left(\left(k_1+k_2\right) \left(k_1+k_2+4 \sigma _1 z_1\right)+4
   \sigma _2 \left(\left(k_1+k_2\right) z_2-\sigma _1
   \left(z_1-z_2\right){}^2\right)\right)\right)}{4 \sigma _2 \sigma
   _3+4 \sigma _1 \left(\sigma _2+\sigma _3\right)}-i \left(\sigma
   _2+\frac{\sigma _1 \sigma _3}{\sigma _1+\sigma _3}\right)
   \xi_2^2}
\\&
    \times
    \psi_1^T(k_1)C\gamma^\mu \xi\!\!/_2n\!\!/\gamma^5\xi\!\!/_2\gamma_\mu\psi_2(k_2)\psi_3(0) .
\end{aligned} 
\end{equation}

The $\widetilde{O}_{a1}$ and $\widetilde{O}_{a2}$ can be simplified into
%\begin{widetext}
\begin{equation}
\begin{aligned}
\widetilde{O}_{a1} = & (d-2)g^2\mu^{4-d}\frac{1}{16\pi^{d/2}}\Gamma(d/2-1)
\int_0^1 d \eta_1\int_0^1 d \eta_2
((z_1-z_2)^2)^{2-d/2}
\\&
\times \psi_1^T((1-\eta_1)z_1+\eta_1z_2))(-Cn\!\!/\gamma^5)\psi_2((1-\eta_2)z_2+\eta_2z_1))\psi_3(0),
\\
\widetilde{O}_{a2} = & g^2\mu^{4-d}\frac{(D-2)^2}{2}\frac{1}{16\pi^{d/2}} \Gamma(-2+d/2)
\int_0^1 d \eta_1\int_0^1 d \eta_2 
((z_1-z_2)^2)^{2-d/2}
\\&
\times \psi_1^T((1-\eta_1)z_1+\eta_1z_2))(-Cn\!\!/\gamma^5)\psi_2((1-\eta_2)z_2+\eta_2z_1))\psi_3(0).
\end{aligned}
\end{equation}

Getting them together:
\begin{equation}
    \widetilde{O}_{a}=
  -\frac{\alpha_s C_F}{4 \pi} 
   \int_0^1 d \eta_1 \int_0^{1-\eta_1} d \eta_2 
   \left( L_{12}^{\text{IR}}-3+\frac{1}{\epsilon_{\mathrm{IR}}}\right)
  \psi_1^{T}\left(z_1\left(1-\eta_1\right)+z_2 \eta_1\right) \widetilde\Gamma \psi_2\left(z_2\left(1-\eta_2\right)+z_1 \eta_2\right) \psi_3(0).
\end{equation}
It is worth noting that when different projectors are chosen for the spatial correlator, only the results of the $\widetilde O_{a1}$ term will be impacted. 
For example, if we switch the vector in $\widetilde \Gamma$ from $n_z^\mu=(0,0,0,1)$ to $n_t^\mu=(1,0,0,0)$, the $\widetilde O_{a1}$ term will acquire an additional negative sign.

%\end{widetext}
\subsubsection{\textbf{Cases~(b) and (c)}}
%\begin{widetext}
For case~(b), after direct calculation, one reaches
%\begin{widetext}
\begin{equation}
\begin{aligned} 
 \widetilde{O}_{b}=&\frac{i(-i)^{3d/2-1}\mu^{4-d}g^2}{16\pi^{d}}
 \int d^d k_1\int d^dk_2\int d^d\xi_2
 \int_0^\infty d \sigma_1 \int_0^\infty d\sigma_2\int_0^\infty d \sigma_3 
 \frac{\sigma_1^{d/2-1}\sigma_2^{d/2-1}\sigma_3^{d/2-2}}{(-i(\sigma_1+\sigma_3))^{d/2}}
\\&
   \times 
   e^{-i \left(\sigma _2+\frac{\sigma
   _1 \sigma _3}{\sigma _1+\sigma _3}\right) \xi_2^2+\frac{i \left(k_1^2 \sigma _2+\sigma _1 \left(4 k_1 \sigma _2
   z_1+k_2^2\right)+\sigma _3 \left(\left(k_1+k_2\right)
   \left(k_1+k_2+4 \sigma _1 z_1\right)-4 \sigma _1 \sigma _2
   z_1^2\right)\right)}{4 \sigma _2 \sigma _3+4 \sigma _1
   \left(\sigma _2+\sigma _3\right)}}
   \\&
   \times
   \psi^T(k_1)C\gamma_\mu(\frac{-k\!\!/_1-2 \left(\sigma _3 \left(\xi\!\!/_2+\frac{k\!\!/_2 \sigma _1+\sigma
   _3 \left(k\!\!/_1+k\!\!/_2+2 \sigma _1 z\!\!/_1\right)}{2 \sigma _2 \sigma _3+2
   \sigma _1 \left(\sigma _2+\sigma _3\right)}\right)+\sigma _1
   z\!\!/_1\right)}{2 \left(\sigma _1+\sigma _3\right)}+z\!\!/_1)n\!\!/_z\gamma_5\psi(z_2) 
   \\&
   \times (\xi\!\!/_2+\frac{k\!\!/_2 \sigma_1+\sigma_3 \left(k\!\!/_1+k\!\!/_2+2 \sigma_1
   z\!\!/_1\right)}{2 \sigma _2 \sigma _3+2 \sigma _1 \left(\sigma
   _2+\sigma _3\right)}\gamma^\mu\psi(k_2)),
\end{aligned}
\end{equation}    
%\end{widetext}
note that the bi-linear part in it is proportional to $p^\mu$ in leading twist, then the spinor terms can be simplified vastly 
%\begin{widetext}
\begin{equation}
\begin{aligned}
\widetilde{O}_{b}= & \frac{(-i)^{d}\mu^{4-d}g^2}{16\pi^{d}}\int d^d k_1\int d^dk_2\int d^d\xi_2
\int_0^\infty d \sigma_1 \int_0^\infty d\sigma_2\int_0^\infty d \sigma_3 
\frac{\sigma_1^{d/2-1}\sigma_2^{d/2-1}\sigma_3^{d/2-1}}{((\sigma_1+\sigma_3))^{d/2+1}}
\\&
\times e^{-i \left(\sigma _2+\frac{\sigma _1 \sigma _3}{\sigma _1+\sigma
_3}\right) \xi_2^2+\frac{i \sigma _1 z_1 \left(k_1 \sigma _2+\sigma _3
   \left(k_1+k_2-\sigma _2 z_1\right)\right)}{\sigma _2 \sigma
   _3+\sigma _1 \left(\sigma _2+\sigma _3\right)}}
%\\&
\psi^T(k_1)C\gamma_\mu\xi\!\!/_2n\!\!/_z\gamma_5\psi(z_2) (\xi\!\!/_2\gamma^\mu\psi(k_2)).
\end{aligned}
\end{equation}    
Then it can be simplified to 
\begin{equation}
\begin{aligned}
\widetilde{O}_{b}=(1-2/d)\frac{d\mu^{4-d}g^2\Gamma(d/2-2)}{32\pi^{d/2}}(z_1^2)^{2-d/2}\int_0^1 d\eta_1 \int_0^1 d \eta_2 
        \psi^T((1-\eta_1)z_1)(-Cn\!\!/_z\gamma_5)\psi(z_2) \psi(\eta_2z_1).
\end{aligned}
\end{equation}    
%\end{widetext}
%\end{widetext}
Following the same routine, all results within this pattern can be written down directly
%these results can be written down directly
%\begin{widetext}
\begin{align}
    \widetilde{O}_{b}= & -\frac{\alpha_s C_F}{8 \pi}  \int_0^1 d \eta_1 \int_0^{1-\eta_1} d \eta_2 
    %\ln \left(\frac{1}{4}\mu^2_{\text{IR}} z_1^2 e^{2 \gamma_E-1}\right)
    \left(L_1^{\text{IR}}-1+\frac{1}{\epsilon_{\mathrm{IR}}}\right)
    \psi_1^{T}\left((1-\eta_1) z_1\right) \widetilde\Gamma \psi_2\left(z_2\right) \psi_3\left(\eta_2 z_1\right) ,
    \\
    \widetilde{O}_{c}= & -\frac{\alpha_s C_F}{8 \pi} \int_0^1 d \eta_1 \int_0^{1-\eta_1} d \eta_2 
    %\ln \left(\frac{1}{4}\mu^2_{\text{IR}} z_2^2 e^{2 \gamma_E-1}\right)
    \left(L_2^{\text{IR}}-1+\frac{1}{\epsilon_{\mathrm{IR}}}\right)
    \psi_1^{T}\left( z_1\right) \widetilde\Gamma \psi_2\left((1-\eta_1)z_2\right) \psi_3\left(\eta_2 z_2\right).
\end{align}
\end{widetext}
It should be noted that case~(a) possesses an additional finite component in comparison to cases~(b) and (c). 
From the calculations presented above, it becomes evident that this difference arises from an extra component in case~(a). 
Notably, there are no terms analogous to those in case (a) in the meson LCDAs calculation.
Additionally, there is no UV divergence in all these operators.

\subsection{W-W pattern}
Last but not least, 
the W-W pattern, corresponding to Fig-~\ref{Pspic} (j, k, l) will be considered.
After the aforementioned preparation, the corresponding one-loop results can be easily obtain
 \begin{align} 
 &\widetilde{O}_{k} = \frac{\alpha _s C_F}{2 \pi }
 %\ln \left(\frac{1}{4}\mu ^2 z_1^2 e^{2 \gamma _{E}+2}\right) 
 \left(L_1^{\text{UV}}+2\right)
 \psi_1^{T}\left( z_1\right) \widetilde\Gamma \psi_2\left(z_2\right) \psi_3(0)  , 
 \\& 
\widetilde{O}_{l} = \frac{\alpha _s C_F}{2 \pi } 
%\ln  \left(\frac{1}{4}\mu ^2 z_2^2 e^{2 \gamma _{E}+2}\right) 
\left(L_2^{\text{UV}}+2\right)
\psi_1^{T}\left( z_1\right) \widetilde\Gamma \psi_2\left(z_2\right) \psi_3(0)  , 
 \\& 
\widetilde{O}_{j} = -\frac{\alpha _s C_F}{4 \pi } 
%\ln  \left(\frac{\mu ^2 \left(z_1^2 z_2^2\right) e^{2 \gamma _{E}+2}}{4 \left(z_1-z_2\right){}^2}\right) 
\left( 
%\left \{\left( \frac{z_1^2 z_2^2}{\left(z_1-z_2\right)^2}\right)^{\text{IR}}\right \}
L_1^{\text{UV}}+L_2^{\text{UV}}-L_{12}^{\text{UV}}
+2\right)
\psi_1^{T}\left( z_1\right) \widetilde\Gamma \psi_2\left(z_2\right) \psi_3(0)  . 
 \end{align}    
Note that in these cases, only UV divergences arise when the two ends of the gluons coincide with each other.
%Only UV divergences arise in these cases.
%And "a" plays the role of a UV regulator.

\begin{widetext}
\subsection{One-loop results for spatial correlator}

%The next-to-leading order Feynman diagrams are shown in Fig.~\ref{Pspic}, and the calculation details are collected in Appendix~\ref{sec:LBDA_one_loop}. 
The final results for the spatial correlator to one-loop order are given as
\begin{align} \label{eq:Mpert}
 &{M}(\nu_1,\nu_2,0,z^2,\mu)=
 % \\& 
  \left\{1 
 + \frac{\alpha _s C_F}{\pi }\left(
 \frac{1}{2}  L_1^\text{UV}
 +
 \frac{1}{2}  L_2^\text{UV}
 +
 \frac{1}{2}  L_{12}^\text{UV}
 +\frac{3}{2}
 \right)\right\} 
 %\\&\times 
 {M}_0 \left(\nu_1,\nu_2,0,z^2,\mu\right) 
 \notag\\& 
%%%%%%%%%%%%%%%%%%%%%%%%%%%%%%%%%%%%%%%%%%%%%
-\frac{\alpha _s C_F}{8 \pi } \int_0^1 d \eta _1 \int_0^{1-\eta _1} d \eta _2 
\left\{ 
2 
\left (L_{12}^\text{IR}-3+\frac{1}{\epsilon_{\mathrm{IR}}}\right )
{M}_0 \left(\left(1-\eta _1\right) \nu_1+\eta _1 \nu_2 , \left(1-\eta _2\right) \nu_2+\eta _2 \nu_1 , 0 , z^2,\mu \right)
\right.
\notag
\\&
\left.
+
\left (L_1^\text{IR}-1+\frac{1}{\epsilon_{\mathrm{IR}}}\right )
{M}_0 \left(\left(1-\eta _1\right) \nu_1 , \nu_2 , \eta _2 \nu_1 , z^2,\mu \right) 
+
\left (L_2^\text{IR}-1+\frac{1}{\epsilon_{\mathrm{IR}}}\right )
{M}_0 \left(\nu_1 , \left(1-\eta _1\right) \nu_2 , \eta _2 \nu_2 , z^2,\mu \right)
\right\}
\notag
 \\& 
 %%%%%%%%%%%%%%%%%%%%%%%%%%%%%%%%%%%%%%%%%
-\frac{\alpha _s C_F}{4 \pi } \int_0^1 d \eta  \left\{
\left(
{M}_0 \left((1-\eta ) z _1 , z _2 , 0 , P^z,\mu \right) 
+
{M}_0 \left(z _1 , z _2 , \eta  z _1 , P^z,\mu \right) 
\right) 
\left\{
\left(
L_1^\text{IR}+1+\frac{1}{\epsilon_{\mathrm{IR}}}
\right)
\left(\frac{1-\eta }{\eta }\right)_+ 
+2 \left(\frac{\ln  \eta  }{\eta }\right)_+ \right\}   
\notag
\right.
\notag 
%%%%%%%%%%%%%%%%%%%%%%%%%%%%5%%%%%%%%%%%%%%%%%%%%%%%%%%%%5
\\& 
+
\left(
{M}_0 \left(z _1 , (1-\eta ) z _2 , 0 , P^z,\mu \right) 
+
{M}_0 \left(z _1 , z _2 , \eta  z _2 , P^z,\mu \right) 
\right)
\left\{
\left(L_2^\text{IR}+1+\frac{1}{\epsilon_{\mathrm{IR}}}
\right)
\left(\frac{1-\eta }{\eta }\right)_+ 
+2 \left(\frac{\ln  \eta  }{\eta }\right)_+ \right\} 
\notag 
%%%%%%%%%%%%%%%%%%%%%%%%%%%%5%%%%%%%%%%%%%%%%%%%%%%%%%%%%5
\\& 
\left.
+
\left(
{M}_0 \left((1-\eta ) \nu_1+\eta \nu_2 , \nu_2 , 0 , z^2,\mu \right)
+{M}_0 \left(z _1 , (1-\eta ) z _2+\eta z _1 , 0 , P^z,\mu \right) 
\right)
\left\{ 
\left (L_{12}^\text{IR}+1+\frac{1}{\epsilon_{\mathrm{IR}}}\right )
\left(\frac{1-\eta }{\eta }\right)_+
+2 \left(\frac{\ln  \eta}{\eta }\right)_+ 
\right\}
\right\},
\end{align}     
\end{widetext}
where $M_0$ stands for tree-level matrix element.
%$z_i^\mu=(0,0,0,z_i)$. 
%To avoid the ambiguity, all the contractions involving $z_i$ will write ``$\cdot$" out explicitly,  and  $k\cdot z_1$, and $z_1 \cdot z_1=-z_1^2$.
We have checked that these results are consistent with the calculation in the momentum space~\cite{Deng:2023csv}. 
Moreover, one can see the UV and IR behaviors clearly in the coordinate space, which is convenient for the renormalization scheme to be established below. 

\subsection{Light-cone correlator}

To obtain the light-cone results, one just needs to perform a similar calculation as the equal-time correlator case, up to  choose $z^2=0$ and $\Gamma=C \gamma_5 \slashed n$.
The light-cone results can be written down straightforwardly as a by-product. 
Besides, they will be used for matching later.
Hence these light-cone results will be outlined.
%At the tree level, the light-cone correlator is: 
%\begin{align}
%&\mathcal{I}_0(z_1,z_2,0,\mu )\equiv \notag
%\\
%&\left\langle 0\left| \psi_1^T(z_1)\Gamma \psi_2(z_2) \psi_3(0)\right|u(x_1 P)d(x_2 P)s(x_3 P)\right\rangle_R,   
%\end{align}
%where $\Gamma=C \gamma_5 \not {n}$ and $n^\mu=(1,0,0,-1)/\sqrt{2}$.
It is important to note that three cases corresponding to the self-energy of Wilson lines do not contribute to the light-cone case.
%will not give any contribution in the light-cone case.
%All cases are depicted in Fig-\ref{LCpic}.
%\begin{figure}[htb]
    %\centering
    %\includegraphics[width=0.5\textwidth]{}
    %\caption{One loop corrections for the light-cone distribution of the $\Lambda$ %baryon.}
 %   \label{LCpic}
%\end{figure}  
%The calculation starts from the LCDA case, which corresponds to the condition $z^2=0$.
For the q-q pattern:
\begin{widetext}
\begin{equation}
\begin{aligned}
O_{a}=&-\frac{ \alpha_s C_F}{4 \pi }
\frac{1}{\epsilon_{\text{IR}}}
\int_0^1 d \eta_1 \int_0^{1-\eta_1} d \eta_2 \psi_1^{T}\left(z_1\left(1-\eta_1\right)+z_2 \eta_1\right) \Gamma \psi_2\left(z_2\left(1-\eta_2\right)+z_1 \eta_2\right) \psi_2(0),
\\
O_{b}= &- \frac{\alpha_s C_F}{8 \pi}
\frac{1}{\epsilon_{\text{IR}}}
\int_0^1 d \eta_1 \int_0^{1-\eta_1} d \eta_2 
\psi_1^{T}\left((1-\eta_1) z_1\right) \Gamma \psi_2\left(z_2\right) \psi_3\left(\eta_2 z_1\right) ,
\\
O_{c}= &- \frac{\alpha_s C_F}{8 \pi} 
\frac{1}{\epsilon_{\text{IR}}}
\int_0^1 d \eta_1 \int_0^{1-\eta_1} d \eta_2 \psi_1^{T}\left( z_1\right) \Gamma \psi_2\left((1-\eta_1)z_2\right) \psi_3\left(\eta_2 z_2\right).
\end{aligned}
\end{equation}    
%\end{widetext}

For the q-W pattern:
\begin{equation}
\begin{aligned}
%\end{aligned}
%\end{equation}
%\begin{widetext}
%\begin{equation}
%	\begin{aligned}
O_{e}=&-\frac{\alpha_s C_F }{2 \pi} \frac{1}{\epsilon_{\text{IR}}}
\int_0^1 d \eta\left(\frac{1-\eta}{\eta}\right)_+
\psi_1^{T}\left((1-\eta) z_1\right) \Gamma \psi_2\left(z_2\right) \psi_3(0) ,
\\
O_{h}=&-\frac{\alpha_s C_F}{4 \pi} 
\frac{1}{\epsilon_{\text{IR}}}
\int_0^1 d \eta\left(\frac{1-\eta}{\eta}\right)_+
\psi_1^{T}\left(z_1\right) \Gamma \psi_2\left(z_2\right) \psi_3\left(\eta z_1\right),
\\
O_{f}=&-\frac{\alpha_s C_F }{2 \pi} 
\frac{1}{\epsilon_{\text{IR}}}
\int_0^1 d \eta\left(\frac{1-\eta}{\eta}\right)_+\psi_1^{T}\left( z_1\right) \Gamma \psi_2\left((1-\eta) z_2\right) \psi_3(0),
\\
O_{i}=&-\frac{\alpha_s C_F}{4 \pi} 
\frac{1}{\epsilon_{\text{IR}}} 
\int_0^1 d \eta\left(\frac{1-\eta}{\eta}\right)_+\psi_1^{T}\left(z_1\right) \Gamma \psi_2\left(z_2\right) \psi_3\left(\eta z_2\right),
\\
O_{d}=&-\frac{\alpha_s C_F }{4 \pi} \frac{1}{\epsilon_{\text{IR}}} 
\int_0^1 d \eta\left(\frac{1-\eta}{\eta}\right)_+
\left(
  \psi_1^T\left((1-\eta) z_1+\eta z_2\right) \Gamma \psi_2\left(z_2\right) \psi_3(0)
  -
  \psi_1^{T}\left((1-\eta) z_1 \right) \Gamma \psi_2\left(z_2\right) \psi_3(0) 
  \right),
    \\
O_{g}=&-\frac{\alpha_s C_F}{4 \pi} \frac{1}{\epsilon_{\text{IR}}}
    \int_0^1 d \eta\left(\frac{1-\eta}{\eta}\right)_+
    %\\&\times 
    \left(
    \psi_1^{T}\left(z_1\right) \Gamma \psi_2\left(\eta z_1+(1-\eta) z_2\right)  \psi_3(0)
    -
    \psi_1^{T}\left(z_1\right) \Gamma \psi_2\left((1-\eta) z_2\right) \psi_3(0)\right) .
\end{aligned}
\end{equation}    
%\end{widetext}

Collecting them together, we have 
%\clearpage
%\begin{widetext}
\begin{equation} 
\begin{aligned} \label{light}
& \mathcal{I}(\nu_1,\nu_2,0,\mu) = \mathcal{I}_0 \left(\nu_1 , \nu_2 , 0 ,\mu \right) 
 \\& 
-\frac{\alpha _s C_F}{8 \pi } \frac{1}{\epsilon_{\text{IR}}}
\left\{ 2 \int_0^1 d \eta _1 \int_0^{1-\eta _1} d \eta _2  \right.
\displaystyle\left[\mathcal{I}_0 \left(\left(1-\eta _1\right) \nu_1+\eta _1 \nu_2 , \left(1-\eta _2\right) \nu_2+\eta _2 \nu_1 , 0 , \mu \right) \right.
\\&
+ \mathcal{I}_0 \left(\left(1-\eta _1\right) \nu_1 , \nu_2 , \eta _2 \nu_1 , \mu \right)
\left.
+ \mathcal{I}_0 \left(z_1 , \left(1-\eta _1\right) \nu_2 , \eta _2 \nu_2 , \mu \right) \right]
\\& 
+2 \int_0^1 d \eta  \left(\frac{1-\eta }{\eta }\right)_+ 
% \\& 
%\times 
\left \{  \left(\mathcal{I}_0 \left((1-\eta ) \nu_1+\eta  \nu_2 , \nu_2 , 0 , \mu \right)+
\mathcal{I}_0 \left(\nu_1 , (1-\eta ) \nu_2+\eta  \nu_1 , 0 , \mu \right) \right) \right.
\\& 
\left.\left.
 + 
 \left(\mathcal{I}_0 \left((1-\eta ) \nu_1 , \nu_2 , 0 , \mu \right)+\mathcal{I}_0 \left(\nu_1 , \nu_2 , \eta  \nu_1 , \mu \right)\right)
+
\left(
\mathcal{I}_0 \left(\nu_1 , (1-\eta ) \nu_2 , 0 , \mu 
\right)
+\mathcal{I}_0 \left(\nu_1 , \nu_2 , \eta  \nu_2 , \mu \right)
\right)
\right \} \right \} . 
\end{aligned} 
\end{equation}    
\end{widetext}
The light-cone correlator can be normalized by dividing it by its 0-momentum matrix element
\begin{equation}
  \mathfrak{I}(\nu_1,\nu_2,\nu_3,\mu) \equiv \frac{\mathcal{I}(\nu_1,\nu_2,\nu_3,\mu)}{\mathcal{I}(0,0,0,\mu)}. 
\end{equation}

\section{Ratio scheme in small spatial separation}\label{sec-ratio}

\subsection{renormalization and Reduced ITD }
In the subsequent sections, all UV divergences will be addressed using the widely adopted ratio scheme. 
Fundamentally, the validity of the ratio scheme relies on the principle of multiplicative renormalization for composite operators. 
When UV renormalization parameters can be factorized out, it becomes feasible to provide a ratio-form definition of the distribution
\begin{equation}
    \mathfrak{M}(\nu_1,\nu_2,\nu_3,z^2)=\left.\displaystyle\frac{M(\nu_1,\nu_2,\nu_3,z^2,\mu)}{M(\omega \nu_1,\omega \nu_2,\omega \nu_3,z^2,\mu)}\right|_{\omega=0}
\end{equation}
%which is named reduced-ITD, 
to cancel UV divergence out.
Based on the one-loop results, it is evident that the ratio scheme will not affect the existing IR physics. 
Additionally, this scheme efficiently eliminates the lattice discreteness effect over short distances. 
Under this scheme, the results of the spatial correlator can be readily converted to
\begin{widetext}
 \begin{align} \label{ratio}
 \mathfrak{M}&(\nu_1,\nu_2,0,z^2)=
 %\\& 
 %\\&\times 
 \mathfrak{M}_0 \left(\nu_1 , \nu_2 , 0 ,z^2\right) 
 \notag\\ 
%%%%%%%%%%%%%%%%%%%%%%%%%%%%%%%%%%%%%%%%%%%%%
-&\frac{\alpha _s C_F}{8 \pi } \int_0^1 d \eta _1 \int_0^{1-\eta _1} d \eta _2   
 \notag\\ 
\times &
\left\{
\left (L_1^\text{IR}-1+\frac{1}{\epsilon_{\mathrm{IR}}}\right )
\left(
\mathfrak{M}_0 \left(\left(1-\eta _1\right) \nu_1 , \nu_2 , \eta _2 \nu_1 ,z^2 \right) 
-
\mathfrak{M}_0 \left(\nu_1 , \nu_2 , 0,z^2 \right)
\right)
\right.
\notag\\&
+
\left (L_2^\text{IR}-1+\frac{1}{\epsilon_{\mathrm{IR}}}\right )
\left(
\mathfrak{M}_0 \left(\nu_1 , \left(1-\eta _1\right) \nu_2 , \eta _2 \nu_2 ,z^2 \right)
-
\mathfrak{M}_0 \left(\nu_1 , \nu_2 , 0,z^2 \right)
\right)
\notag\\&
\left.
+2 
\left (L_{12}^\text{IR}-3+\frac{1}{\epsilon_{\mathrm{IR}}}\right )
\left(
\mathfrak{M}_0 \left(\left(1-\eta _1\right) \nu_1+\eta _1 \nu_2 , \left(1-\eta _2\right) \nu_2+\eta _2 \nu_1 , 0 ,z^2\right) 
-
\mathfrak{M}_0 \left( \nu_1,\nu_2 , 0 ,z^2 \right)
\right)
\right\}
 \notag\\ 
 %%%%%%%%%%%%%%%%%%%%%%%%%%%%%%%%%%%%%%%
 -&\frac{\alpha _s C_F}{4 \pi } \int_0^1 d \eta 
 \\
 \times &\left\{
 \left(
 \mathfrak{M}_0 \left(\nu_1 , (1-\eta ) \nu_2 , 0 ,z^2 \right)
 +
 \mathfrak{M}_0 \left(\nu_1 , \nu_2 , \eta  \nu_2 ,z^2 \right)
\right)
\left\{
\left( L_2^\text{IR}+1+\frac{1}{\epsilon_{\mathrm{IR}}}
\right)
\left(\frac{1-\eta }{\eta }\right)_+ 
+2 \left(\frac{\ln  \eta  }{\eta }\right)_+ \right\}  
\right.
%%%%%%%%%%%%%%%%%%%%%%%%%%
\notag\\& 
+\left(
\mathfrak{M}_0 \left((1-\eta ) \nu_1 , \nu_2 , 0 ,z^2 \right)
+\mathfrak{M}_0 \left(\nu_1 , \nu_2 , \eta  \nu_1 ,z^2 \right) 
\right)
\left\{ 
\left (L_1^\text{IR}+1+\frac{1}{\epsilon_{\mathrm{IR}}}\right )
\left(\frac{1-\eta }{\eta }\right)_+
+2 \left(\frac{\ln  \eta }{\eta }\right)_+\right\} 
 \notag\\ &
\left.
+\left(
\mathfrak{M}_0 \left((1-\eta ) \nu_1+\eta \nu_2 , \nu_2 , 0 ,z^2 \right)
+
\mathfrak{M}_0 \left(\nu_1 , (1-\eta ) \nu_2+\eta \nu_1 , 0 ,z^2 \right)
\right)
\left\{
\left (L_{12}^\text{IR}+1+\frac{1}{\epsilon_{\mathrm{IR}}}\right )
\left(\frac{1-\eta }{\eta }\right)_+ 
+2 \left(\frac{\ln  \eta }{\eta }\right)_+ \right\}
\right\}.
\notag
\end{align}       
\end{widetext}
It should be mentioned that the validity of multiplicative renormalization has not been proved. 
General proof to all orders is still required. Furthermore, it can be verified through lattice computations, which will help clarify the range of its validity.

\subsection{matching and RGE}
Although the light-cone correlator and spatial correlator are both ITD with different coordinate choices.
But one can not simply turn the spatial correlator to the light-cone correlator by approaching the $z^2$ to zero due to the divergences.

The connection between equal-time and light-cone physics can be established through a matching process:
\begin{align}
&\mathfrak{M}\left(\nu_1,\nu_2,0,z^2\right) =\int_{0}^1 \mathrm{~d} \eta_1 \int_{0}^{1-\eta_1} \mathrm{~d} \eta_2 
\\&
\times C\left(\eta_1,\eta_2,\nu_1,\nu_2, \mu\right) \mathcal{I}\left(\eta_1,\eta_2,\nu_1,\nu_2, \mu\right)+\mathcal{O}\left(z^2\right),
\end{align}
where $C\left(\eta_1,\eta_2,\nu_1,\nu_2, \mu\right)$ stands for the matching kernel.
It should be mentioned that due to the complex nature of the dependence on $\eta_i$s and $\nu_i$s, we present the arguments in a non-standard form within $\mathcal{I}$.

Combining Eq.~\ref{light} and Eq.~\ref{ratio}, we have  
%These matching relationships can be determined by employing the light-cone Operator Product Expansion (OPE)\cite{Anikin:1978tj,Balitsky:1987bk}:
\begin{widetext}
 \begin{equation} 
 \begin{aligned}  \label{matching}
& \mathfrak{M}(\nu_1,\nu_2,0,z_i^2)=
%%%%%%%%%%%%%%%%%%%%%%%%%%%%%%%%%%%%%%%%%%%%%
\mathfrak{I}(\nu_1,\nu_2,0,\mu )-\frac{\alpha _s C_F}{8 \pi } \int_0^1 d \eta _1 \int_0^{1-\eta _1} d \eta _2  
 \\& 
\times \left\{ 
\left (L_1^\text{UV}-1\right )
\mathfrak{I} \left(\left(1-\eta _1\right) \nu_1 , \nu_2 , \eta _2 \nu_1 ,\mu \right) 
+
\left (L_2^\text{UV}-1\right )
\mathfrak{I} \left(\nu_1 , \left(1-\eta _1\right) \nu_2 , \eta _2 \nu_2 ,\mu \right)
\right.
\\&
+2 
\left (L_{12}^\text{UV}-3\right )
\mathfrak{I} \left(\left(1-\eta _1\right) \nu_1+\eta _1 \nu_2 , \left(1-\eta _2\right) \nu_2+\eta _2 \nu_1 , 0  ,\mu \right) 
\\&
\left.
-
\left (L_1^\text{UV}-1\right )
\mathfrak{I} \left(\nu_1 , \nu_2 , 0,\mu  \right) 
-
\left (L_2^\text{UV}-1\right )
\mathfrak{I} \left(\nu_1 ,  \nu_2 , 0,\mu  \right)
-2 
\left (L_{12}^\text{UV}-3\right )
\mathfrak{I} \left(\nu_1,\nu_2, 0 ,\mu  \right) 
\right\}
 \\& 
 %%%%%%%%%%%%%%%%%%%%%%%%%%%%%%%%%%%%%%%%%
-
\frac{\alpha _s C_F}{4 \pi } \int_0^1 d \eta  
% \\& 
\left\{
%%%%%%%%%%%%%%%%%%%%%%%%%%
\left(
\mathfrak{I} \left((1-\eta ) \nu_1 , \nu_2 , 0 ,\mu  \right)
+
\mathfrak{I} \left(\nu_1 , \nu_2 , \eta  \nu_1 ,\mu  \right) 
\right)
    \left\{ 
    \left (L_1^\text{UV}+1\right )
    \left(\frac{1-\eta }{\eta }\right)_+
    +2 \left(\frac{\ln  \eta }{\eta }\right)_+
    \right\} 
    \right.
\\ &
 +\left(
\mathfrak{I} \left(\nu_1 , (1-\eta ) \nu_2 , 0 ,\mu  \right)
+\mathfrak{I} \left(\nu_1 , \nu_2 , \eta  \nu_2 ,\mu  \right) 
\right)
    \left\{ 
    \left (L_2^\text{UV}+1\right )
    \left(\frac{1-\eta }{\eta }\right)_+
    +2 \left(\frac{\ln  \eta }{\eta }\right)_+
    \right\}
\\&
\left.
+\left(
\mathfrak{I} \left(\nu_1 , (1-\eta ) \nu_2+\eta \nu_1 , 0 ,\mu \right)+\mathfrak{I} \left(\nu_1 , (1-\eta ) \nu_2+\eta \nu_1 , 0 ,\mu 
\right) 
\right)
\left\{ 
\left (L_{12}^\text{UV}+1\right )
\left(\frac{1-\eta }{\eta }\right)_+
+2 \left(\frac{\ln  \eta}{\eta }\right)_+ 
\right\} 
\right\}.
\end{aligned} 
\end{equation}    
\end{widetext}
As expect, the IR structures within the light-cone operator and spatial correlator are identical and cancel each other out in the calculation.
Once have extracted the light-cone correlator from the spatial correlator using Eq.~\ref{matching}, one can immediately obtain the LCDAs through performing Fourier transformation.
However, given that the ratio scheme is merely valid in the perturbative region, other methods need to be invoked in other regions.
In regions beyond the perturbative region, it is possible to utilize various models or engage in direct global fitting procedures.
For example, in paper \cite{Han:2023xbl}, the long-distance physics are addressed by the self-renormalization.

%It should be emphasized that this relation is only valid for small $z^2$.
%Upon establishing this connection, the Light-cone correlator can be directly extracted from lattice data.

Moreover, the evolution equation can be derived. 
By following the standard procedure for constructing Renormalization Group Equations (RGE), one can deduce
\begin{widetext}
\begin{equation}
\begin{aligned}
&\mu \frac{\partial}{\partial \mu} 
\mathfrak{I}\left(\nu_1, \nu_2, 0, \mu\right)=
\\&
\frac{\alpha_s C_F}{4 \pi} \int_0^1 d \eta_1 \int_0^{1-\eta_1} d \eta_2 
%\\
%& \times 
\left\{
2 \mathfrak{I}\left((1-\eta_1) \nu_1+\eta_1 \nu_2,\left(1-\eta_2\right) \nu_2+\eta_2 \nu_1, 0, \mu\right)
-
2 \mathfrak{I}\left(\nu_1, \nu_2, 0,\mu\right)
\right. 
\\&
\left.
+
\mathfrak{I}\left(\left(1-\eta_1\right) \nu_1, \nu_2, \eta_2 \nu_1,\mu\right) 
-
\mathfrak{I}\left(\nu_1, \nu_2, 0, \mu\right) 
+
\mathfrak{I}\left(\nu_1,\left(1-\eta_1\right) \nu_2, \eta_2 \nu_2, \mu\right)
-
\mathfrak{I}\left(\nu_1, \nu_2,0, \mu\right)
\right\}
\\&
+\frac{\alpha_s C_F}{2 \pi} \int_0^1 d \eta\left(\frac{1-\eta}{\eta}\right)_{+} 
%\\& \times
\left\{
\mathfrak{I}\left((1-\eta) \nu_1+\eta \nu_2, \nu_2, 0, \mu\right)\right. 
 +
\mathfrak{I}\left(\nu_1,(1-\eta) \nu_2+\eta \nu_1, 0, \mu\right) 
\\
&\left. +
\mathfrak{I}\left((1-\eta) \nu_1, \nu_2, 0, \mu\right) 
 + 
\mathfrak{I}\left(\nu_1,(1-\eta) \nu_2, 0, \mu\right)
 %\\& 
 +
\mathfrak{I}\left(\nu_1, \nu_2, \eta \nu_1, \mu\right) 
 +
\mathfrak{I}\left(\nu_1, \nu_2, \eta \nu_2, \mu\right) \right\},
\end{aligned}
\end{equation}    
\end{widetext}
which is consistent with the discussions made in paper \cite{Braun:1999te}.
It is worth noting that unlike the cases of meson distribution amplitudes (DAs) or parton distribution functions (PDFs), where the Renormalization Group Equation (RGE) can be obtained by taking derivatives with respect to $\ln z^2$. 
This can be substantiated by the following considerations.
For meson DAs or PDFs, there are only one variables need to be consider, the $\ln z^2$ used for taking derivative is clear.
While in the baryon case, there are three distinct $\ln z^2$ dependencies, namely $\ln z^2_1$, $\ln z^2_2$, and $\ln (z_1-z_2)^2$, need to be considered, which makes it hard to perform the derivative.
Even if we were to focus on derivatives involving just one of the $\ln z^2$, it remains difficult to obtain a kernel-like result in a concise form.
In meson cases, the use of translation invariance allows for a shift in the dependence on integral variables, resulting in a more compact expression.
%, namely the kernel. 
%In the baryon case, the presence of translation invariance among the three fields complicates the shift operation and prevents us from achieving a similarly concise form for the involved integrals.
In the baryon case, the translation invariance extends to three fields. 
When trying to perform a shift operation similar to what's done in meson cases, it always includes an additional quark. 
This makes it challenging to express the results in a more concise manner.
Therefore, the equivalence between the RGE and derivatives with respect to $\ln z^2$ of the spatial correlator are not explicit in this context.

\section{Summary}
To obtain the light baryon LCDAs through lattice QCD using spatial correlation, we have calculated the spatial correlator to one-loop order. 
And we have conducted a comprehensive analysis of their ultraviolet (UV) and infrared (IR) properties.
Subsequently, we applied renormalization via the ratio scheme.
The renormalized spatial correlator has been related to the light baryon correlator through matching.
%Through which the light baryon correlator can be obtained directly. 
This allows for the direct extraction of the light-baryon light-cone correlator.
These results provide a fundamental methodology for extracting the baryon LCDA from first principles. 
The validation of the ratio renormalization scheme for the light baryons is based on the multiplicative renormalization, which still need a robust proof in the future studies.
%The ratio scheme can be utilized in the short distance efficiently.
%By cooperating with other models or engaging in parameter fitting for long distances, we can obtain meaningful results.
The efficiency and utility of the adopted ratio scheme have been demonstrated, offering a practical approach to investigate the light baryon LCDA on the lattice.
This procedure can also be extended for the examination of other light-cone physical quantities of light baryons in collaboration with lattice QCD in future research endeavors.

%过去：背景
%历史
%实验
%理论
%现在：当前工作
%理论，所用方法
%所得结果，和之前进行比较
%将来：
%理论的进一步发展
%实验证实预言

%%%%%%%%%%%%%%%%%%%%%%%%%%%%%%%%%%%%%%%%%%%%%%%%%%%%%%%%%%%%%
%%%%%%%%%%%%%%%%%%%%%%%%%%%%table %%%%%%%%%%%%%%%%%%%%%%%%
%%%%%%%%%%%%%%%%%%%%%%%%%%%%%%%%%%%%%%%%%%%%%%%%%%%%%%%%%%%%

\begin{acknowledgments}
We would like to thank Wei Wang, Yushan Su, Jun Zeng, and Zhifu Deng for their insightful comments and invaluable discussions.
This work is supported in part by the Natural Science Foundation of China under Grants No.12125503, No. U2032102, No. 12061131006,  and No. 12335003.
%We acknowledge support from the National Natural Science Foundation of China (No. 11875306) and the Key Research Program of the Chinese Academy of Sciences (No. XDPB15).
\end{acknowledgments}
\appendix
% Produces the bibliography via BibTeX.
%\bibliographystyle{JHEP}
%\bibliography{DHB}

\bibliography{ratio}

\end{document}